\documentclass{article}
\usepackage[preprint]{spconf}

\usepackage[utf8]{inputenc}
\usepackage{enumitem} % enumerate
\usepackage[acronym]{glossaries} % this needs to be one of the first packages, otherwise hyperlinks ?
\usepackage{cite}
\usepackage[hyphens]{url}
\usepackage[hidelinks,draft]{hyperref}
\usepackage{graphicx}
\usepackage{amsmath,amssymb,amsfonts}
\usepackage{multirow,makecell}
\usepackage{subcaption}
\usepackage{tikz,tikzscale,xcolor}
\usepackage{algorithmic,textcomp}
\usepackage{flushend} % automagically balances columns in the last page
\usepackage{eso-pic}

\graphicspath{{figures/}}

\newcommand{\ac}[1]{\gls{#1}}
\newcommand{\acp}[1]{\glspl{#1}}

	\newacronym{ads}{ADS}{Absolute Difference of Sums}
	\newacronym{afd}{AFD}{Absolute First Difference}
	\newacronym{amvp}{AMVP}{Advanced Motion Vector Prediction}
	\newacronym{asic}{ASIC}{Application Specific Integrated Circuit}
	\newacronym{avc}{AVC}{Advanced Video Coding}

	\newacronym{bd}{BD}{Bj{\o}ntegaard Delta}
	\newacronym{bdbr}{BD-Rate}{Bj{\o}ntegaard Delta Bitrate}
	\newacronym{bdpsnr}{BD-PSNR}{Bj{\o}ntegaard Delta PSNR}
	\newacronym{bma}{BMA}{Block Matching Algorithm}
	\newacronym{bpp}{bpp}{bits per pixel}

    \newacronym{cabac}{CABAC}{Context-Adaptive Binary Arithmetic Coding}
	\newacronym{cb}{CB}{Coding Block}
	\newacronym{cfm}{CFM}{Coded block Fast Mode}
	\newacronym{cnn}{CNN}{Convolutional Neural Network}
	\newacronym{codec}{CODEC}{Coder/decoder}
	\newacronym{cpu}{CPU}{Central Processing Unit}
	\newacronym{ctb}{CTB}{Coding Tree Block}
	\newacronym{ctc}{CTC}{Common Test Conditions}
	\newacronym{ctu}{CTU}{Coding Tree Unit}
	\newacronym{cu}{CU}{Coding Unit}

	\newacronym{dag}{DAG}{Direct Acyclic Graph}
	\newacronym{dct}{DCT}{Discrete Cosine Transform}
	\newacronym{de}{DE}{Disparity Estimation}
	\newacronym{dlt}{DLT}{Divisible Load Theory}
	\newacronym{dpcm}{DPCM}{Differential Pulse-Code Modulation}
	\newacronym{dram}{DRAM}{Dynamic Random Access Memory}
	\newacronym{ds}{DS}{Diamond Search}
	\newacronym{dsp}{DSP}{Digital Signal Processing}
	\newacronym{dst}{DST}{Discrete Sine Transform}

	\newacronym{ecu}{ECU}{Early CU}
	\newacronym{esa}{ESA}{Successive Elimination Exhaustive Search Algorithm}
	\newacronym{esd}{ESD}{Early Skip Detection}

	\newacronym{fbma}{FBMA}{Fullsearch Block Matching Algorithm}
	\newacronym{fme}{FME}{Fractional Motion Estimation}
	\newacronym{fbsme}{FBSME}{Fixed Block-Size Motion Estimation}
	\newacronym{fht}{FHT}{Fast Hadamard Transform}
	\newacronym{fpga}{FPGA}{Field-Programmable Gate Array}
	\newacronym{fps}{fps}{frames per second}
	\newacronym{fs}{FS}{Fullsearch}
	\newacronym{fsm}{FSM}{Finite State Machine}

	\newacronym{gea}{GEA}{Global Elimination Algorithm}
	\newacronym{gop}{GOP}{Group of Pictures}
	\newacronym{gpp}{GPP}{General Purpose Processor}
	\newacronym{gpu}{GPU}{Graphics Processing Unit}

	\newacronym{hd}{HD}{High Definition}
	\newacronym{hevc}{HEVC}{High Efficiency Video Coding}
	\newacronym{hm}{HM}{HEVC Test Model}
	\newacronym{hsad}{HSAD}{Hadamard Transformed coefficients of residual signal}
	\newacronym{ht}{HT}{Hadamard Transform}

	\newacronym{ime}{IME}{Integer Motion Estimation}
	\newacronym{irs}{IRS}{Iterative Random Search}
	\newacronym{iso}{ISO}{International Standardization Organization}
	\newacronym{itut}{ITU-T}{International Telecommunication Union -- Telecommunication Standardization Sector}

    \newacronym{jem}{JEM}{Joint Exploration Model}
	\newacronym{jm}{JM}{Joint Model}
	\newacronym{jvet}{JVET}{Joint Video Exploration Team}
	\newacronym{jvt}{JVT}{Joint	Video Team}

    \newacronym{klt}{KLT}{Karhunen-Loève Transform}

	\newacronym{lb}{LB}{Linear Buffer}
	\newacronym{lcb}{LCB}{Largest Coding Block}
	\newacronym{ldp}{LDP}{Low Delay P}
	\newacronym{lsb}{LSB}{Least Significant Bit}
	\newacronym{lh}{LH}{Low-Vdd/High-Vt}
	\newacronym{lstm}{LSTM}{Long Short-Term Memory}

    \newacronym{mad}{MAD}{Mean Absolute Deviation}
	\newacronym{mb}{MB}{Macroblock}
	\newacronym{mc}{MC}{Motion Compensation}
	\newacronym{mcm}{MCM}{Multiplierless Constant Multiplication}
	\newacronym{md}{MD}{Mode Decision}
	\newacronym{me}{ME}{Motion Estimation}
	\newacronym{mlp}{MLP}{Multilayer Perceptron}
	\newacronym{mse}{MSE}{Mean Squared Error}
	\newacronym{mv}{MV}{Motion Vector}
	\newacronym{mvp}{MVP}{Motion Vector Predictor}

	\newacronym{nn}{NN}{Neural Network}

	\newacronym{pb}{PB}{Prediction Block}
	\newacronym{pde}{PDE}{Partial Distortion Elimination}
	\newacronym{pmd}{PMD}{Portable Mobile Device}
	\newacronym{pmv}{PMV}{Predicted Motion Vector}
	\newacronym{poc}{POC}{Picture Order Count}
	\newacronym{psad}{PSAD}{Partial SAD}
	\newacronym{psatd}{PSATD}{Partial SATD}
	\newacronym{psnr}{PSNR}{Peak signal-to-noise ratio}
	\newacronym{pu}{PU}{Prediction Unit}

	\newacronym{qme}{QME}{Quartet-pel motion estimation}
	\newacronym{qp}{QP}{Quantization Parameter}
	\newacronym{qt}{QT}{Quadtree}
	\newacronym{qtbt}{QTBT}{Quadtree Binary-Tree}
	\newacronym{qtmt}{QTMT}{Quadtree Multi-type Tree}

	\newacronym{ra}{RA}{Random Access}
	\newacronym{rbsad}{RBSAD}{Reduced Bit Sum of Absolute Differences}
	\newacronym{rd}{RD}{Rate-Distortion}
	\newacronym{rdo}{RDO}{Rate-Dis\-tor\-tion Optimization}
	\newacronym{relu}{ReLU}{Rectified Linear Unit}
	\newacronym{rtl}{RTL}{Register Transfer Level}

	\newacronym{sad}{SAD}{Sum of Absolute Differences}
	\newacronym{sat}{SAT}{Summed-Area Table}
	\newacronym{satd}{SATD}{Sum of Absolute Transformed Differences}
	\newacronym{scb}{SCB}{Smallest Coding Block}
	\newacronym{sdc}{DC$^{\text{\textregistered}}$}{Synopsys$^{\text{\textregistered}}$ Design Compiler}
	\newacronym{sea}{SEA}{Successive Elimination Algorithm}
	\newacronym{sic}{ICC$^{\text{\textregistered}}$}{Synopsys$^{\text{\textregistered}}$ IC Compiler}
	\newacronym[plural=SRAMs,firstplural=Static Random Access Memories (SRAMs)]{sram}{SRAM}{Static Random Access Memory}
	\newacronym[plural=SPMs,firstplural=Scratchpad Memories (SPMs)]{spm}{SPM}{Scratchpad Memory}
	\newacronym{ssd}{SSD}{Sum of Squared Differences}
	\newacronym{sse}{SSE}{Sum of Squared Errors}
	\newacronym{svcs}{VCS$^{\text{\textregistered}}$}{Synopsys$^{\text{\textregistered}}$ Verilog Compiler Simulator}
	\newacronym{svm}{SVM}{Support-Vector Machine}

	\newacronym{tb}{TB}{Transpose Buffer}
	\newacronym{tsmc}{TSMC}{Taiwan Semiconductor Manufacturing Company Limited}
	\newacronym{tss}{TSS}{Three-Step Search}
	\newacronym{tu}{TU}{Transform Unit}
	\newacronym{tzs}{TZS}{Test Zone Search}

	\newacronym{uhd}{UHD}{Ultra High Definition}
	\newacronym{umh}{UMH}{Uneven MultiHexagon}
	\newacronym{umhs}{UMHS}{Unsymmetrical Cross Algorithm Grid Search}

	\newacronym{vbs}{VBS}{Variable Block Size}
	\newacronym{vbsme}{VBSME}{Variable Block Size Motion Estimation}
	\newacronym{vlsi}{VLSI}{Very-large-scale Integration}
	\newacronym{vtm}{VTM}{VVC Test Model}
	\newacronym{vvc}{VVC}{Versatile Video Coding}

	\newacronym{wht}{WHT}{Walsh-Hadamard Transform}
	\newacronym{wpp}{WPP}{Wavefront Parallel Processing}

\hypersetup{breaklinks=true}

\usetikzlibrary{calc}

\copyrightnotice{\copyright\ 2021 IEEE}

\toappear{To appear in {\it Proc.\ ICASSP 2021, June 06-11, 2021, Toronto, Canada}}

\title{Relying on a rate constraint to reduce Motion Estimation complexity}

\name{
	Gabriel B. Sant'Anna, Luiz Henrique Cancellier, Ismael Seidel, Mateus Grellert, Jos\'{e} Lu\'{i}s G\"{u}ntzel
	\thanks{
		This study was financed in part by the Coordenação de Aperfeiçoamento de Pessoal de Nível Superior - Brasil (CAPES) - Finance Code 001 and by Brazilian Council for Scientific and Technological Development (CNPq) through Project Universal (437924/2018-1) and PQ grant (312077/2018-1).
	}
}

\address{
	Embedded Computing Laboratory (ECL) - Dept. of Computer Science and Statistics (INE)\\
	Federal University of Santa Catarina (UFSC) - Florian\'{o}polis, Brazil\\
	\{baiocchi.gabriel, luizhenriquecancellier, ismaelseidel\}@gmail.com,\\ \{mateus.grellert, j.guntzel\}@ufsc.br
}

\begin{document}
% \ninept

\AddToShipoutPictureBG*{
    \begin{tikzpicture}[remember picture,overlay]
        \node[align=justify,text width=\paperwidth * 0.90,draw,line width=1pt,fill=yellow] at ($(current page.north) + (0,-1.75cm)$) {\footnotesize
        \copyright 2021 IEEE.  Personal use of this material is permitted.  Permission from IEEE must be obtained for all other uses, in any current or future media, including reprinting/republishing this material for advertising or promotional purposes, creating new collective works, for resale or redistribution to servers or lists, or reuse of any copyrighted component of this work in other works.};
    \end{tikzpicture}
}

\begingroup
\let\newpage\relax
\maketitle
\endgroup

\begin{abstract} % 100 to 150 words
This paper proposes a rate-based candidate elimination strategy for Motion Estimation, which is considered one of the main sources of encoder complexity.
We build from findings of previous works that show that selected motion vectors are generally near the predictor to propose a solution that uses the motion vector bitrate to constrain the candidate search to a subset of the original search window, resulting in less distortion computations.
The proposed method is not tied to a particular search pattern, which makes it applicable to several ME strategies.
The technique was tested in the VVC reference software implementation and showed complexity reductions of over 80\% at the cost of an average 0.74\% increase in BD-Rate with respect to the original TZ Search algorithm in the LDP configuration.
\end{abstract}

\begin{keywords} % max 5
Video Coding, Integer Motion Estimation, Block-Matching Algorithm, Rate-Constraint, VVC
\end{keywords}

\section{Introduction}

	Due to a growing consumer demand for video playback services and high-definition content, video data amounts to the majority of global internet traffic \cite{Cisco} -- even more so during the COVID-19 pandemic, which has caused a worldwide increase in digital media consumption \cite{Statista}.
	As the demand for transmitted video content expands, the need for advanced compression techniques increases.

	The Joint Video Exploration Team of ITU-T developed the \ac{vvc}, a state-of-the-art video coding standard improving over the previous one, the \ac{hevc}.
	The newly achieved compression ratio is paired with a rise in encoder complexity: the \ac{vvc} reference software implementation presents an average 44.4\% bitrate reduction at the cost of an approximately 10$\times$ higher encoding time w.r.t. that of \ac{hevc} \cite{Comparison}.
	This increased complexity prevents some applications -- such as real-time streaming and video coding in battery-constrained devices -- from making full use of the new standard, thus motivating research efforts to improve encoding performance.

	\ac{me} has been regarded as one of the most time-consuming operations of video compression, being responsible for 40\% to 80\% of encoding time when using \ac{fs} \cite{ME}.
	Many different fast \ac{me} algorithms attempt to reduce this cost by limiting the \ac{ime} search region to as few points as possible \cite{Chessboard}, but while this strategy does improve performance, the search becomes sub-optimal and may get trapped in a local minimum \cite{Diamond}.
	The \ac{tzs} algorithm, adopted as the default \ac{ime} method in both of the reference software implementations of \ac{hevc} \cite{HM} and \ac{vvc} \cite{VTM}, mitigates this problem by combining a variant of Diamond Search with the semi-exhaustive Raster Search \cite{HardTZS}.

	Although \ac{tzs} brings \ac{me} complexity down when compared to \ac{fs}, the number of searched points is still considered too large for real-time applications \cite{ME} \cite{Classification}.
	In addition, the iterative nature of its search steps prevents efficient parallel implementations \cite{HardTZS}.
	The ``Octagonal-Axis Raster'' search, proposed by \cite{Octagonal}, exemplifies how a smaller number of search points can reduce \ac{me} complexity with a negligible increase in \ac{bdbr}.
	It was designed to account for \ac{mv} distributions averaged over decoded video sequences in order to limit the search to regions with the highest occurrences of the best \acp{mv}.
	Unfortunately, this method cannot be generalized for different algorithms because it operates in a specific \ac{tzs} step.
	Meanwhile, the work in \cite{Order} leveraged the assumption that the likelihood of finding optimal candidates decreases as the \ac{mv} bitrate increases to produce a rate-ordered variant of Successive Elimination \cite{Sea}.
	Their algorithm provides the optimal solution (in terms of coding efficiency) within the search set, but limits the complexity reduction that can be achieved.

	In light of the observations brought up by \cite{Octagonal} and \cite{Order}, that selected \ac{mv} distribution and \ac{mv} bitrate surfaces should be taken into account when designing a \ac{me} search pattern, we explore the impact of the \ac{mv} bitrate estimation on fast \ac{ime} algorithms.
	Thus, our contributions are:
	\begin{enumerate}[wide, labelindent=0pt]
		\item We bring evidence that previous \ac{ime} search patterns are related to \ac{mv} bitrate.
		To the best of our knowledge, no previous work explicitly explores this relation \cite{ME} to define the search pattern, but those patterns frequently seem to match the estimated bitrate cost surface;
		\item A strategy to reduce \ac{ime} complexity by explicitly using the bitrate component of the cost function as a criterion to eliminate distortion calculations on a per-candidate basis.
		Unlike existing fixed search patterns, our technique is flexible enough to be used in conjunction with other fast algorithms and also allows for different complexity reduction and quality targets by parameterizing the elimination threshold;
		\item An evaluation and discussion considering \ac{ldp} and \ac{ra} configurations, combining our strategy with the \ac{tzs} implementation of the \ac{vtm} as a case study.
	 \end{enumerate}

	This paper is organised as follows: we highlight key concepts of the \ac{me} process in Section \ref{sec:background} and then explain the proposed algorithm in Section \ref{sec:algorithm}.
	Section \ref{sec:experiments} details our experiments and displays obtained results, which are followed by a brief discussion and possible future works in Section \ref{sec:conclusions}.

\section{Background}\label{sec:background}

	\ac{me} has three distinct steps: \ac{mv} Prediction, Integer \ac{me} and Fractional \ac{me} \cite{ME}.
	\ac{mv} Prediction uses the motion information of neighboring blocks to derive a \ac{mvp}, which defines the starting position of the subsequent search.
	Then, \ac{ime} searches a region centered around the \ac{mvp} for a \ac{mv} which minimizes the cost function.
	At last, Fractional \ac{me} performs a refinement process over the integer result.
	We focus on the second step, namely \ac{ime}.

	Equation \eqref{cost} shows the cost function minimized during \ac{me} \cite{RDO}, where $r$ estimates the bitrate of the difference between a given \ac{mv} ($\vec{mv}$) and the \ac{mvp} ($\vec{mvp}$), $\lambda$ is the Lagrange multiplier increasing the weight of the bitrate component and $d$ computes distortion between the ``original'' pixel block ($\mathbf{O}$) given as input to the \ac{me} process and a ``candidate'' block ($\mathbf{C}^{\vec{mv}}$), which is offset to the original block by the given \ac{mv}.
	\begin{equation}
		j(\vec{mv}) = d(\mathbf{C}^{\vec{mv}}) + \lambda \cdot r(\vec{mv} - \vec{mvp})
		\label{cost}
	\end{equation}

	The \ac{sad}, shown in \eqref{distortion}, is commonly used as distortion metric due to its simplicity.
	However, its calculation is needed for every evaluated candidate, requiring large sample blocks to be loaded from memory, making it the most time-consuming computation in \ac{ime}.
	\begin{equation}
		d(\mathbf{C}) = \sum\limits_{i = 1}^m {\sum\limits_{j = 1}^n { |\mathbf{C}_{i,j} - \mathbf{O}_{i,j}| }}
		\label{distortion}
	\end{equation}

\begin{figure}[t]
		\centering
		\includegraphics[width=\columnwidth,keepaspectratio]{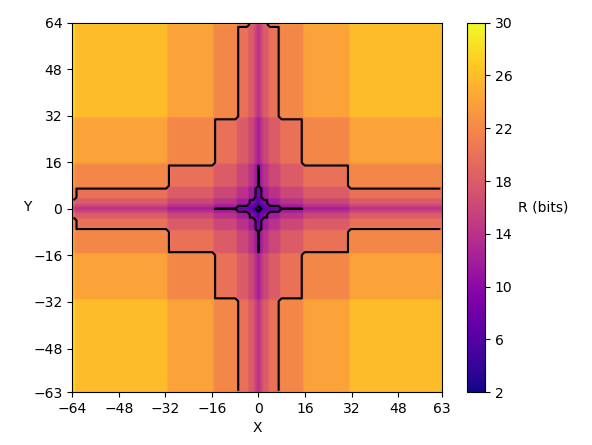}
		\caption{Bitrate surface for \ac{mv} coordinates relative to the \ac{mvp}. Black lines highlight bitrate values of 20 (biggest shape), 10 (cross-shaped contour) and 4 (small diamond).}
		\label{fig:rate-surface}
		\vspace{-0.25cm}
	\end{figure}

	The bitrate estimation function, in turn, is implemented in the \ac{vtm} as \eqref{rate}, where $g(v)$ denotes the length of a signed Exponential Golomb code for an integer value and each vector component is given in coordinates relative to the \ac{mvp}.
	Fig. \ref{fig:rate-surface} displays the bitrate surface of a search region of $128 \times 128$ pixels centered around the \ac{mvp}.
	We highlight that $r(\vec{mvd})$ can be efficiently implemented because it does not depend on the video's content and $g(v)$ can be calculated with a lookup table since the search window size limits the range of possible values for $\vec{mvd}_x$ and $\vec{mvd}_y$ \cite{RateSea}.
	\begin{equation}
		r(\vec{mvd}) = g(\vec{mvd}_x) + g(\vec{mvd}_y)
		\label{rate}
	\end{equation}

\begin{figure}[h]
		\centering
		\includegraphics[width=\columnwidth,keepaspectratio]{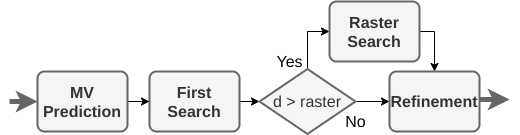}
		\caption{High-level flowchart of the \ac{tzs} algorithm.}
		\label{fig:tzs_flow}
		\vspace{-0.25cm}
	\end{figure}

	Fig. \ref{fig:tzs_flow} shows a flowchart of \ac{tzs}, which is the default \ac{ime} algorithm in \ac{vtm}.
	Its first step chooses a starting point for the following search, similarly to \ac{mv} Prediction but with added \acp{mv}.
	Around the starting \ac{mv}, the First Search step expands a diamond.
	When the First Step best candidate is found at a distance $d$ greater than the raster step size, a Raster Search is performed to find a closer match.
	Finally, the Refinement Step iteratively expands a diamond centered in the current best candidate's position until it either finds a fixed point or makes a maximum number of attempts, in which case it returns the current best candidate.

\section{Rate-based Candidate Elimination}\label{sec:algorithm}

	Although the \ac{tzs} algorithm significantly reduces complexity when compared to an exhaustive search, related works show that it is still possible to narrow down the number of evaluated candidates with only a small loss in coding efficiency.
	Notably, previous works regarding \ac{me} in \ac{hevc} have shown that selected \acp{mv} are generally located around the predictor: it has been stated that 87\% of them can be found after the \ac{tzs} prediction step \cite{eTZS} and that over 94\% of the best \acp{mv} are within a small, diamond-shaped range around the \ac{mvp} \cite{Classification}.

\begin{figure}[t]
		\centering
		\includegraphics[width=\columnwidth,keepaspectratio]{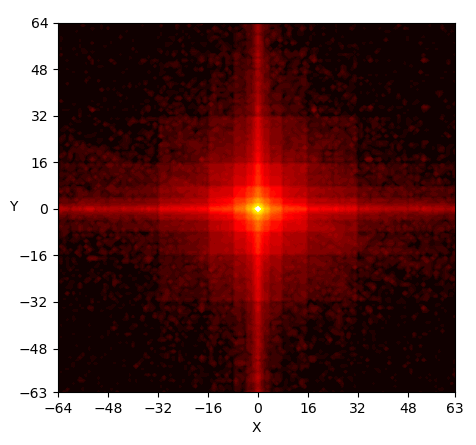}
		\caption{Selected \ac{mv} distribution averaged over the set of tested sequences (see Section \ref{sec:experiments}). Brighter colors represent an exponentially higher number of decisions at a given point.}
		\label{fig:mv-heatmap}
		\vspace{-0.25cm}
	\end{figure}

	Fig. \ref{fig:mv-heatmap} shows a heatmap with the spatial distribution of \acp{mv} chosen by the \ac{vtm} encoder.
	When we create points by element-wise pairing the values from Fig. \ref{fig:rate-surface} and the log values of Fig. \ref{fig:mv-heatmap}, a Pearson correlation coefficient of $-0.89$ can be found, showing that the number of selected \acp{mv} exponentially decreases as the bitrate estimate increases.
	This allows us to conclude that most decisions are likely within a fraction of the search region where estimated bitrate values are smaller.

	Therefore, our approach consists in reducing the average complexity of the cost function $j(\vec{mv})$ computation by skipping distortion computations for all \acp{mv} for which $r(\vec{mv} - \vec{mvp}) > t$, where $t$ is a threshold value.
	Three reasons led us to using this approach:
	1) the function can be efficiently calculated with lookup tables, so skipped blocks do not need to be fetched from memory;
	2) there is a direct correlation between effectively selected candidate distributions and the vector rate surface;
	3) the criterion can be applied on top of existing \ac{ime} algorithms, and the threshold can be adjusted to suit a specific application's constraints.
	A possible disadvantage would be its reliance on the \ac{mv} prediction step efficacy.

	When using this technique, the \ac{ime} search window is expected to be constrained to a diamond-shaped region centered around the \ac{mvp}, which is extended over its axes and has a radius proportional to the threshold $t$.
	For instance, when this candidate elimination strategy is used with $t=20$, the search region has a similar shape to the one proposed in \cite{Octagonal}, but our strategy is applied to the entire \ac{tzs} execution instead of only to the Raster step.
	In the end, although not able to guarantee the cost function minimization, the rate-constrained algorithm should manage to eliminate most block distortion computations while still evaluating the regions more likely to contain optimal candidates.

\section{Results and Discussion}\label{sec:experiments}

	In order to evaluate the elimination criterion proposed in the previous section, the \ac{tzs} algorithm implemented in \ac{vtm} 6.2 was modified\footnote{Code is available at \\ \url{https://gitlab.com/baioc/vtm/tree/rate-elimination}} to apply rate-based elimination.
	The experiments were performed using \ac{ldp} and \ac{ra} configurations, setting \textit{InternalBitDepth} to 8 instead of its default of 10.
	The video sequences analyzed are the \ac{ctc} subset common to both \ac{hevc} \cite{CTC-HEVC} and \ac{vvc} \cite{CTC-VVC}, and each sequence was encoded using \acp{qp} 22, 27, 32 and 37 in order to evaluate coding efficiency using the \ac{bdbr} metric \cite{Bjontegaard}.

	In this paper, the complexity $C$ of a video is defined as:
	\begin{equation}
		C = \sum_{s \in S} totalCandidates(s) \times area(s)
		\label{complexity}
	\end{equation}
	In \eqref{complexity}, $S$ is the set of all possible \ac{cu} sizes in \ac{vvc}, $totalCandidates(s)$ represents the total number of candidates with size $s$ for which distortion was calculated during \ac{ime} and $area(s)$ is the area of that \ac{cu} size.
	Considering the original \ac{vtm} \ac{ime} complexity as $C_{ori}$ and the \ac{ime} complexity of a modified implementation as $C_{mod}$, the complexity reduction $\Delta C$ is calculated as follows:
	\begin{equation}
		\Delta C = \frac{C_{ori} - C_{mod}}{C_{ori}} \times 100\%
	\end{equation}
	This complexity metric, unlike time measurements, is not affected by compiler optimizations or machine specifications and thus allows for reproducible experiments and fair comparisons with other \ac{ime} search algorithms.

	\begin{figure}[t]
		\centering
		\includegraphics[width=\columnwidth]{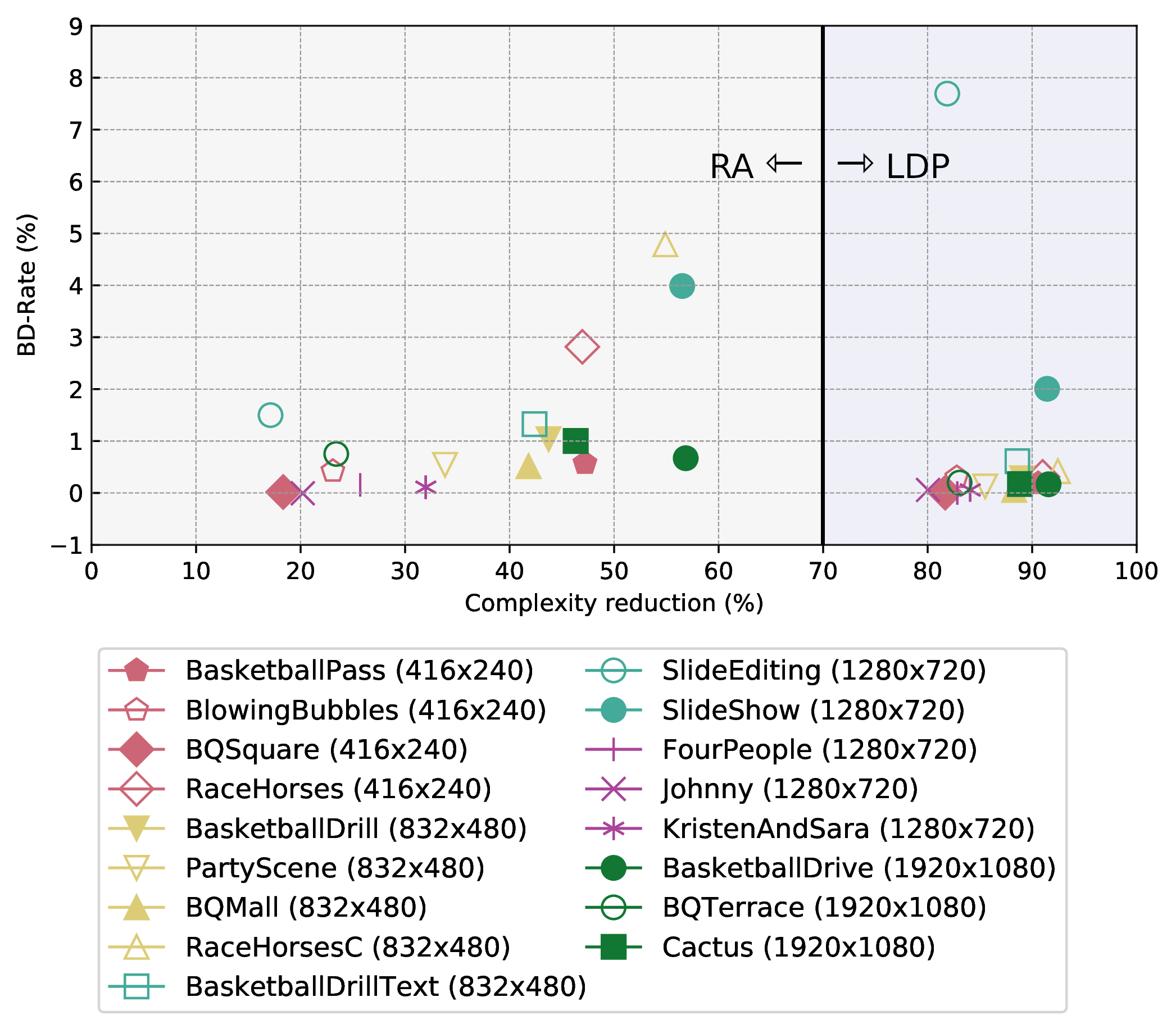}
		\caption{Rate-based candidate elimination results using the original \ac{tzs} algorithm as a baseline. The chart includes all 17 tested sequences for both \ac{ra} and \ac{ldp} configurations.}
		\label{fig:results-pruning}
		\vspace{-0.30cm}
	\end{figure}

	We initially fixed the elimination threshold at $t=4$ as this is the smallest value for which \ac{tzs} will test candidates other than the \ac{mvp}.
	Fig. \ref{fig:results-pruning} shows our results.
	Looking at \ac{ra}, even though most test sequences have \ac{bdbr} increases of less than 1\%, these results show that $t=4$ is too restrictive in some cases.
	Therefore, given that a larger search region should result in lower \acp{bdbr} as it eliminates less candidates, we conducted additional experiments with the sequences that presented \ac{bdbr} values above 1\% while relaxing the elimination criterion by using $t$ equal to 10 and 20.
	Notably, $t=20$ allows the effective search window (see Fig. \ref{fig:rate-surface}) to approximate that of the Octagonal-axis pattern presented in \cite{Octagonal}, which is known to have near zero \ac{bdbr} increases.
	Thus, with the purpose of comparing our results with those of \cite{Octagonal}, we have replicated their search pattern -- originally implemented for \ac{hevc} -- in \ac{vtm}, adapting it to any search window size and non-square dimensions.

	Table \ref{tab:results-octagonal} shows the results obtained when increasing the elimination threshold in the \ac{ra} configuration and compares them with those of the Octagonal-axis raster pattern.
	Applying $t=10$ suffices to decrease \ac{bdbr} levels of three sequences to less than 1\%, maintaining complexity reduction significantly higher than that of Octagonal-axis.
	For the other videos, $t=20$ is required to further reduce \ac{bdbr} and approximate the results of Octagonal-axis, while still having better complexity results.

	\begin{table}[b]
		\caption{Complexity reduction (\%) and \ac{bdbr} increase (\%) when using different thresholds in the \ac{ra} configuration.}
		\centering
		\resizebox{\linewidth}{!}{
		\begin{tabular}{|l|c|c|c|c|c|c|}
			\hline
			\multirow{2}{*}{Sequence} & \multicolumn{2}{c|}{$t=10$} & \multicolumn{2}{c|}{$t=20$} & \multicolumn{2}{c|}{Octagonal-axis} \\ \cline{2-7}
			              & BDBR & $\Delta C$ & BDBR & $\Delta C$ & BDBR  & $\Delta C$ \\ \hline
			Cactus        & 0.55 & 39.0       & 0.12 & 28.8       & 0.02  & 26.2 \\ \hline
			BballDrill    & 0.60 & 35.8       & 0.14 & 23.4       & 0.01  & 22.2 \\ \hline
			BballDrillTxt & 0.67 & 34.3       & 0.09 & 22.9       & -0.04 & 21.2 \\ \hline
			SlideEdit     & 1.23 & 10.0       & 0.63 &  6.8       & 0.03  &  6.1 \\ \hline
			RHorses       & 1.53 & 38.0       & 0.35 & 25.1       & 0.02  & 22.5 \\ \hline
			SShow         & 2.88 & 51.0       & 0.88 & 42.5       & -0.05 & 36.4 \\ \hline
			RHorsesC      & 3.02 & 47.7       & 0.70 & 33.9       & 0.10  & 30.2 \\ \hline
		\end{tabular}
		}
		\label{tab:results-octagonal}
	\end{table}

	Meanwhile, the \ac{ldp} configuration shows promising results, with an overall \ac{bdbr} below 1\% for $t=4$, strengthening the premise that good \ac{mv} candidates can generally be found within the vicinity of the \ac{mvp}.
	Table \ref{tab:results-ldp} shows complexity reduction and \ac{bdbr} results for \ac{ldp}, averaged by class.
	While the Octagonal-axis search reduces 16.4\% of the \ac{ime} complexity in the best case, our approach achieves complexity reductions of over 80\% with negligible \ac{bdbr} increase in most cases -- the two screen-content sequences in class F being the only exceptions.

	\begin{table}[t]
		\caption{Complexity reduction (\%) and \ac{bdbr} increase (\%) per-class averages in the \ac{ldp} configuration.}
		\centering
		\begin{tabular}{|c|c|c|c|c|}
			\hline
			\multirow{2}{*}{Class} & \multicolumn{2}{c|}{$t=4$} & \multicolumn{2}{c|}{Octagonal-axis} \\ \cline{2-5}
			   & BDBR & $\Delta C$ & BDBR & $\Delta C$ \\\hline
			B  & 0.18 & 87.8       & 0.02  & 13.9 \\ \hline
			C  & 0.22 & 88.8       & 0.00  & 15.2 \\ \hline
			D  & 0.20 & 86.5       & 0.04  & 10.7 \\ \hline
			E  & 0.04 & 82.3       & -0.04 &  6.5 \\ \hline
			F  & 3.44 & 87.3       & 0.37  & 16.4 \\ \hline
		\end{tabular}
		\label{tab:results-ldp}
		\vspace{-0.2cm}
	\end{table}

\section{Conclusions}\label{sec:conclusions}

	In this paper, we have showed how \ac{mv} bitrate can influence \ac{ime} search patterns and proposed an algorithm that explicitly uses this relation to prune search regions through an efficient and simple criterion which is applied on a per-block basis.
	The main advantages of our proposal are its versatility, being possible to combine this solution with existing \ac{ime} search algorithms; and its low complexity, given that the bitrate estimation function can be easily computed with lookup tables.

	Our experiments show that even when rate-based elimination is applied on top of \ac{tzs} with a fixed threshold of 4, we are able to reduce \ac{ime} complexity by approximately 86.69\% when using the \ac{ldp} configuration, with a small (0.74\% average \ac{bdbr}) coding efficiency loss.
	Although \ac{ra} configuration results displayed a lower complexity reduction and a higher coding efficiency loss when using the same threshold, additional experiments showed that changing the threshold to 20 allows \ac{bdbr} increase to be kept below 1\%, while still able to reduce complexity slightly further than the Octagonal-axis raster search pattern.
	This exemplifies how the algorithm can be configured for different trade-offs between complexity reduction and encoding efficiency.

	We reckon future works could propose modifications and improvements over this technique, as well as study its use on top of other fast algorithms.
	For instance, the identification of specific corner case test sequences with increased \ac{bdbr} points towards adaptive thresholds to accommodate content-dependent motion characteristics.
	Finally, since a small threshold value was able to produce satisfying results for the \ac{ldp} configuration, hardware-accelerated implementations could apply a static elimination criterion (\textit{e.g.} over \ac{fs}), to drastically simplify the \ac{ime} search, reducing it to a small region around the predictor in order to obtain significant reductions to circuit area, power and cost.

\bibliographystyle{IEEEbib}
\bibliography{ms}

\end{document}